# Characterisation, performance, and operational aspects of the H4RG-15 near infrared detectors for the MOONS instrument

Derek Ives*, Domingo Alvarez, Naidu Bezawada, Elizabeth George, and Benoit Serra
European Southern Observatory, Karl-Schwarzschild-Strasse 2, 85748 Garching bei München, Germany

## ABSTRACT

MOONS is a multi-object spectrograph for the ESO VLT covering a simultaneous wavelength range of 0.6-1.8 microns using approximately 1000 fibres. It uses four Teledyne Imaging Systems H4RG-15 4K x 4K detectors with 2.5 µm cut-off material for the two longer wavebands (YJ and H). Since the spectrographs utilize an extremely fast modified Schmidt camera design, then the detectors are situated in the optical beam and hence required the development of a novel 64-channel cryogenic differential cryogenic preamplifier, minimized for optical footprint which will be reported on.

We have operated the Engineering Grade H4RG detector at a range of temperatures from 90K to 40K and we will report on the advantages of the lower operational temperature. We have completed a full persistence analysis and are able to model it reasonably well and will offer a correction for it in our data pipeline. We have also configured the detector for use with both unbuffered and buffered outputs and report on the differences in performance between the two output types. We have also seen some programming issues with the detector type and will report on our work-around for this, we will also describe the use of the column-deselect feature and row skipping to minimize the effects of PEDs. We have also measured charge injection per read and report on this, likewise we have also measured the inter-pixel capacitance in different regions of the detector. We will present all these results together with a summary of the complete performance characteristics of this detector family.

## 1. INTRODUCTION

MOONS[1] is a multi-object spectrograph that will be mounted at a Nasmyth focus at the VLT. It will have ~1000 fibers deployable over a field of view of ~500 square arcmin. The total wavelength coverage is from 0.6 micron to 1.8 micron, and it will have a low resolution and a medium resolution mode. In the medium resolution mode (R~4,000-6,000) the entire wavelength range 0.65-1.8 micron can be observed simultaneously, while the high resolution mode covers simultaneously two selected spectral regions: one around the CaII triplet (at R~9,000) to measure radial velocities and a second at R~20,000 in the H-band, for detailed measurements of chemical abundances.

Two spectrographs are required due to the large number of fibres. A single cryostat will contain both of them, in order to reduce the overall mass and volume of the instrument. Each spectrograph will simultaneously obtain spectra in the three wavebands, namely RI, YJ and H. A pair of dichroic filters is used to split the light into the three bands. In each band the light will be dispersed and then imaged through a camera onto its associated detector. Two 4kx4k CCD detectors, LBNL fully depleted devices, will be used for the shortest waveband (RI), whilst four Teledyne H4RG-15 detectors will be used for the two longer wavebands (YJ and H).

In the course of the development of the IR detector systems for MOONS, the Engineering Grade (EG) H4RG has gone through an extensive period of testing, optimization, and characterisation, before we do the final installation of the Science Grade (SG) devices into the instrument. All of the usual detector characteristics have been measured but we would like to highlight the comparison in device performance between buffered and unbuffered operation. The device has also been characterised at different operating temperatures from 90K to 40K and we report on the advantages of operation at specific temperatures.

*dives@eso.org

This paper summarizes some of the results obtained in the characterisation of this EG device, but more importantly disseminates what we have learned about the performance and operational aspects of the H4RG that differ from ESO's extensive experience with the H2RG detectors and this will be useful for the future projects using H4RG detectors.

With delivery of the MOONS, HARMONI[2], and MICADO[3] instruments, ESO will procure a total of 23 science-grade H4RG-15 detectors in the next few years so a future paper will report on the performance characteristics seen across so many devices.

## 2. MOONS H4RG NEAR INFRARED DETECTORS

The H4RG-15 is the next generation IR detector for infrared instrumentation with a very large (4096 x 4096) pixel array width and 15 μm pixel pitch. Each detector is more than 61 mm square in size, an astonishing achievement in hybridised technology, even more so when one considers that the devices are manufactured to a flatness better than 20 μm peak-to-valley. Our devices are substrate removed, with 2.5 μm cut-off HgCdTe and will operate using the full 64 outputs. They have been specified to have performance specifications similar in most aspects to the mature H2RG detectors with some loosening of the cosmetic specifications in the corners of the devices. For MOONS we have under-sized the optical beam so that the corners are not used.

### 2.1 Detector operation and requirements

The MOONS instrument has a novel operational concept in which large surveys will comprise the bulk of the observations performed with the instrument, allowing a fairly constant operation mode of the detectors. The typical spectroscopic exposures will be hundreds of seconds long, with sky and object fibres projected onto the same detector. This requires low noise and high dynamic range to both avoid saturation of bright objects and record high signal-to-noise spectra of the sky and any fainter objects. Additionally, it is critical that sky and object spectra can be matched, which requires stable detector operation with no crosstalk between channels. We have focused our H4RG detector system development on meeting these operational requirements with the best performance possible.

## 3. CRYOGENIC PREAMPLIFIER

To achieve the very wide optical field, MOONS is based on a modified Schmidt camera concept which means that the H4RG detector is situated in the centre of the light path. We do not use the SIDECAR ASIC but our own inhouse electronics which uses a cryogenic preamplifier close to the detector. This is required to be positioned entirely behind the detector and with a smaller footprint than the detector to minimize optical vignetting. This was an extreme challenge during the development phase of the design.

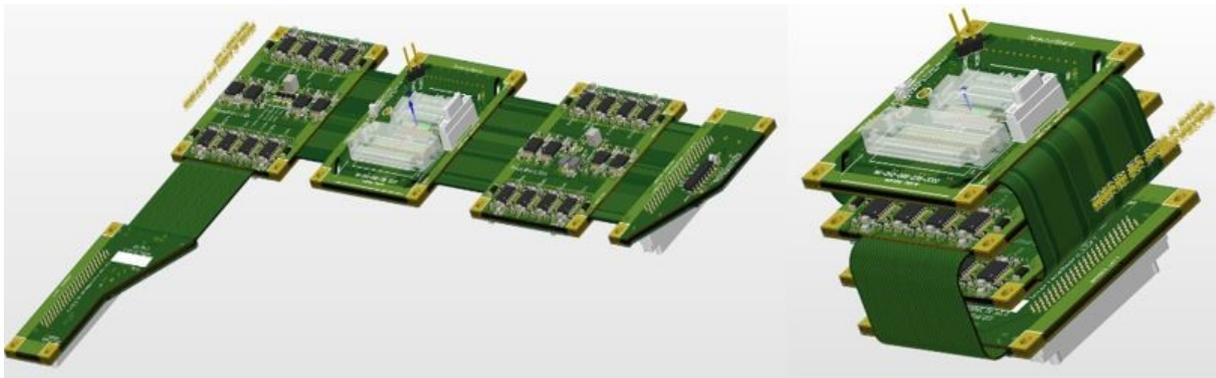

Figure 1, Cryogenic preamplifier rigid flexi PCB unfolded (left) and after folding (right).

In order to achieve the reduced footprint, the preamp has been developed using a Rigid-Flex PCB design and designed to ensure an area of only 50 x 50 mm of rigid PCB area. There was no possibility to include all 64 video differential channels plus clocks, biases, signal conditioning and interfaces (connectors) into a single PCB piece. Therefore, the design is an origami of folding technology using the flexi-rigid PCB as shown in Figure 1 which shows a 3D CAD model of the preamp in its flat configuration on the left and then the nominal setup is given to the right after the origami folding. The cryogenic preamplifier is based on a low noise, low power, low input offset voltage and ultra low drift quad CMOS op-amp, part number OPA4192, which has been shown to operate at <90K with a 500kHz pixel rates in our pseudo-differential configuration. This op-amp is based on CMOS technology and is manufactured using a e-trimmed process to minimize the errors associated with the input offset voltages and temperature drifts. It has a gain bandwidth of 10MHz with a sufficient slew rate (20V/μs) for our needs and a voltage noise of 5.5 nV√Hz. In addition to its excellent DC precision and AC performance, the device offers a high capacitive load drive capability of up to 1nF, which makes it attractive for driving signals over long cable lengths for slow speed operation of the detectors as is the case for our MOONS instrument. All the op-amps we use for cryogenic operation have recently been tested to compare their performance between room temperature and cryogenic operation at 90K, this includes frequency response, noise, d.c. drift and output impedance. These results will be published at a later stage. Figure 2 shows the backside of the preamplifier on the left and the right side image indicates how it interfaces to the detector.

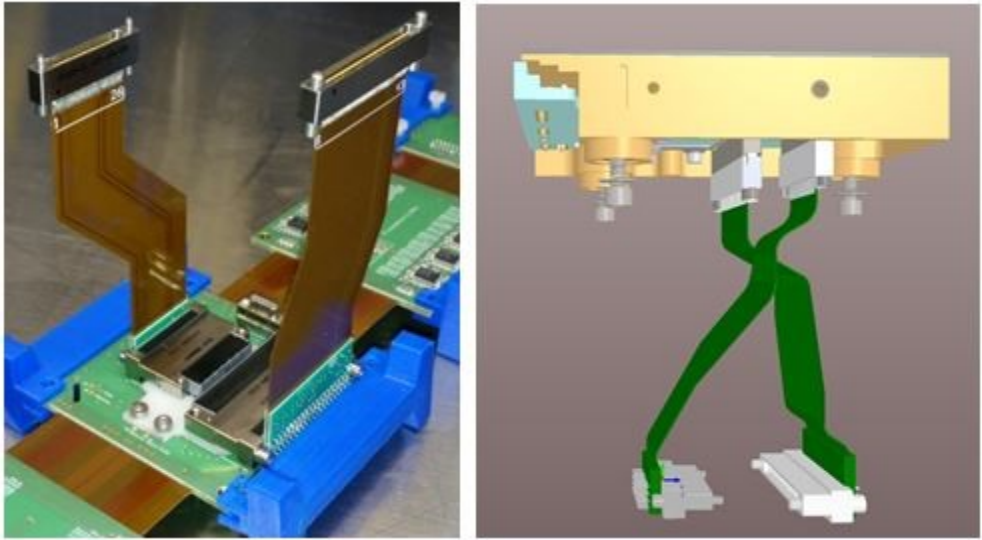

Figure 2, backside of cryogenic preamplifier showing flexis to detector (left) and the detector mount model indicating entrance of the same flexis to the detector (right).

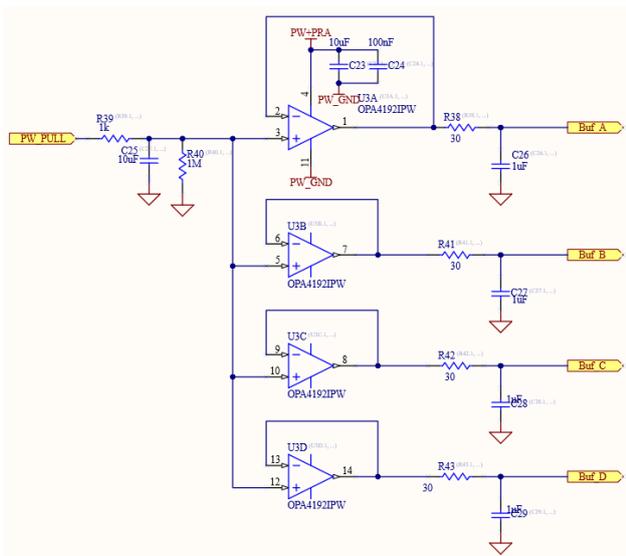

Figure 3, schematic of buffered output load supply circuit.

The preamplifier offers all 64 video channels in both a buffered and unbuffered configuration. For buffered operation then the detector *Vdda* bias is fed via individual op-amps and associated resistive loads to each individual buffered detector output. Figure 3 shows the schematic for our buffered supply topology. In the schematic the "PW_PULL" signal is the detector *Vdda* bias to the detector. Here it goes to the quad op-amp package to generate a low impedance supply per output, in this case for 4 outputs. Each of the supply lines "Buf_A" etc then connect via a resistor direct to the detector buffered output pin. The detector has been tested in both the buffered and unbuffered setup and the differences in performance observed is described later in this paper. The end of the differential video circuit is the ubiquitous NGC controller which has been the ESO standard controller for many years.

# 4. BUFFERED VERSUS UNBUFFERED DETECTOR OPERATION

The H4RG detector has 64 outputs which can each be configured for "unbuffered" mode where the pixel source follower, which has quite high output impedance, drives the signal directly off chip. It can also be configured in a "buffered" mode where another on-chip source follower is switched in, which offers lower output impedance but which requires an external resistive or constant current load. In the past, ESO has always operated their detectors in the unbuffered mode to simplify the electronics and give best noise performance but has recently developed the electronics to operate the H4RG detector in buffered mode to determine the performance advantages.

The buffered mode should have lower output impedance which helps to minimize crosstalk between channels both at the detector and on the cryogenic preamplifier PCB. However, the unknown was the noise performance with the extra buffer stage and load. The electrical difference between the two modes is shown in the H4RG manual for interested readers. We use the PMOS FET buffered option with the pull-up to the *Vdda* bias and use a separate low impedance supply for each output to minimize crosstalk as already described.

## 4.1 Unbuffered bias optimisation

Most of our early days' detector characterisation was carried out unbuffered with 32 outputs. In this normal mode we had some issues trying to get the unit cells to operate at the required pixel rate of 100 kHz. It was eventually possible, but we had to drop the *Vbiasgate* voltage to 1.7V, from the nominal value of 2.3V to increase the unit cell current enough to allow for this pixel speed. We determined optimal operation by doing frequency response measurements using a flat field and reset window and determined the positive and negative going signal swing response at the either edge of the window. However, we discovered that there are detrimental consequences to operating the detector with such a low *Vbiasgate* voltage in that we saw a gain change or "wobble" in the photon transfer plot. This is clearly seen in the plot of Figure 4. On the left the unit cell current is set correctly but the diode bias uses standard TIS values. On the right, it has been optimized with clearly better performance. The bias across the diodes needed adjustment by moving them more positive to ensure a good photon transfer. We attribute this issue to the increased current and low *Vbiagate* voltage setting affecting the normal operating point of the FET in the unit cell.

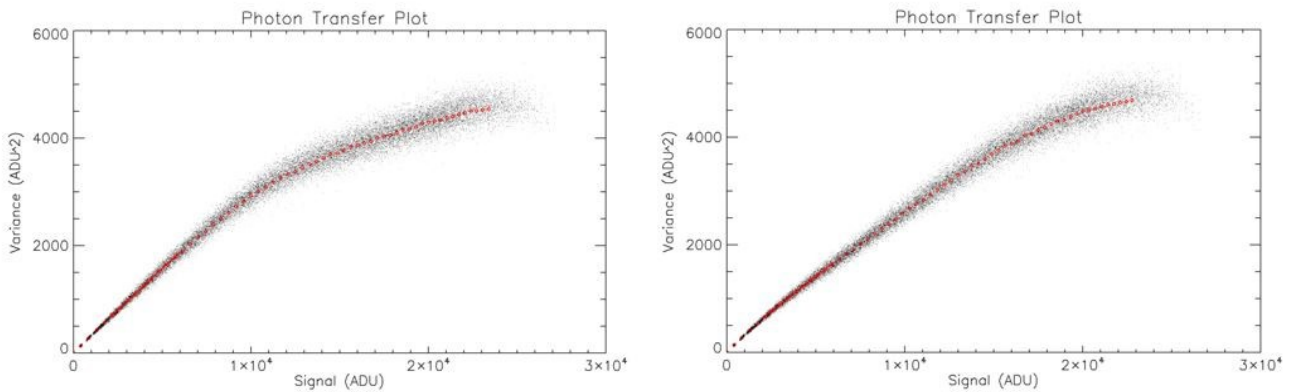

Figure 4, unbuffered output PTC plot with standard TIS bias settings (left) and with ESO optimised bias settings (right).

## 4.2 Comparison of noise performance buffered versus unbuffered

Our testing showed that the buffered outputs give the same noise performance as for the unbuffered outputs. The unbuffered noise performance is typically what we see in H2RG detectors, approximately 13 e- rms for a CDS with the EG device. The buffered outputs should allow the detector to be able to run at a higher pixel rate than the unbuffered output and indeed, this is the case where for the H4RG we can operate faster than 300 kpixel/second for the buffered outputs compared to 100 kpixel/second for the unbuffered outputs. With 64 outputs configured means that we can read the full 16 million pixels out in approximately 1 second. For this paper most of the detector characterization is done with a 200 kHz pixel rate and a 1.5 second frame time. To confirm the noise performance of the buffered outputs we did long Up-The-Ramp (UTR) exposures for both buffered and unbuffered configurations and compared the read noise between both. The result of this test is clearly seen in Figure 5, in the first plot we show that the noise versus number of UTR frames is the same for both buffered and unbuffered, the extra source follower stage does not add any extra noise to the video chain. In the second plot we see the advantage of the higher frame rate in that we can achieve the lowest achievable UTR noise in a shorter timescale, meaning we can achieve optimal SNR performance, faster.

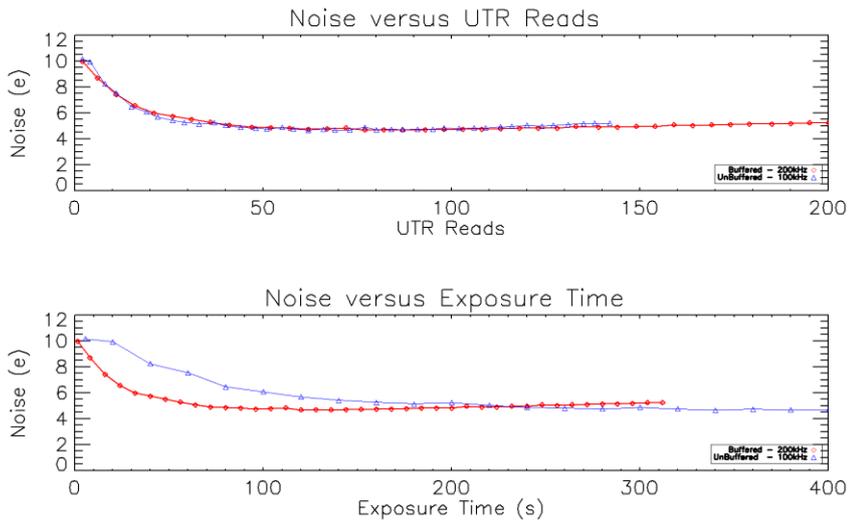

Figure 5, noise versus UTR reads (top) and noise versus exposure time (bottom) using same data set, note red plot is for buffered outputs and blue plot is for unbuffered outputs.

### 4.3 AC and DC crosstalk between output channels

A hoped for advantage of the buffered outputs was to reduce the crosstalk seen between channels on the detector. Typically, this crosstalk can be capacitive[4] and/or resistive. The capacitive coupling is typically seen as spikes in each output associated with a real change in signal seen in one output. It is normally associated with capacitive coupling between the high impedance outputs of the detector and other tracking on the PCB close to the detector. The resistive crosstalk is due to IR voltage drops on the detector itself or in the dc bias supplies to the detector. The crosstalk is reduced in both cases by employing the lower impedance buffered outputs which helps to minimize the capacitive coupling between tracks and also by having separate low impedance supplies for each buffered output on the detector *Vdaa* bias which helps to reduce any current droop on this output bias, as described earlier in this paper. To test for the AC crosstalk, we use the method described in [8] in which we hold a windowed region which is positioned on one output in reset and then read the whole detector out looking for coupled signals to other outputs. We repeat this process for all the used outputs of the detector. This mean 32 outputs for our unbuffered configuration and 64 outputs for the buffered configuration. We can then plot a 2D matrix which shows the cross talk from any one channel to any other. This is shown in Figure 6 for the buffered case on the left and for the unbuffered case on the right. For the unbuffered case we see that there is indeed crosstalk at the 0.1% level between many outputs and others which are not even nearest neighbours. This implies that the crosstalk is capacitive coupling between the high impedance tracking of the unbuffered output which runs close to other output lines on our cryogenic preamplifier. For the buffered case the output impedance is much lower reducing the coupling to almost nothing. The one column issue seen in the buffered measurement is not crosstalk but an actual column defect on the device which seems to affect some of the outputs.

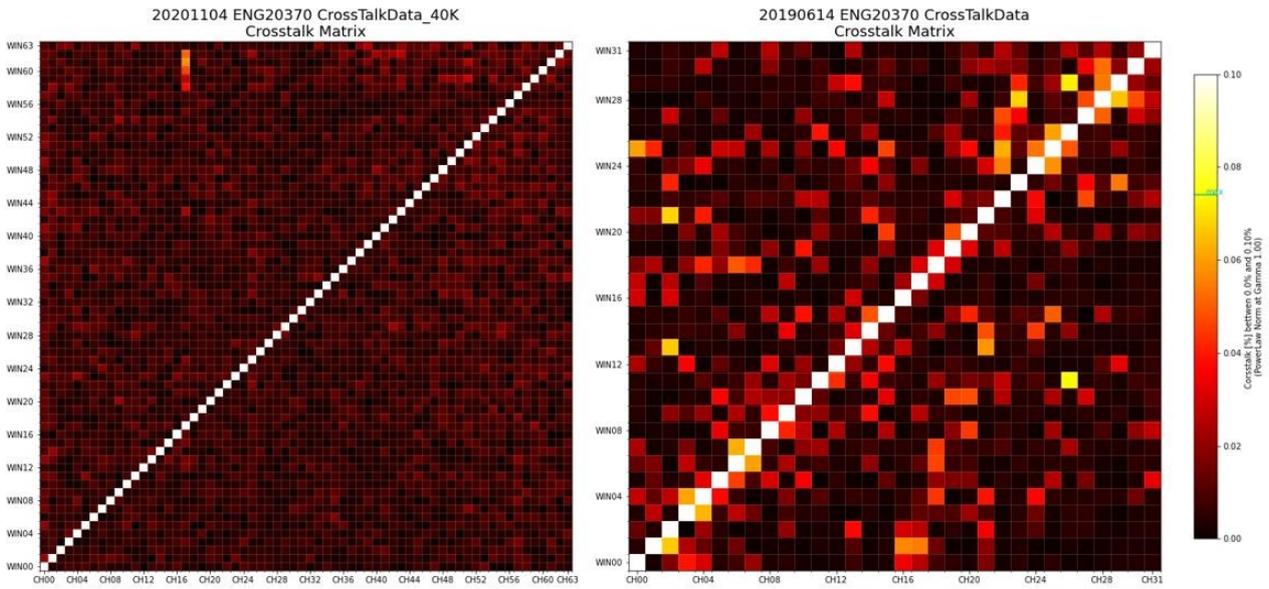

Figure 6, AC crosstalk matrices for H4RG detector, left side for 64 buffered outputs and right side for 32 unbuffered outputs.

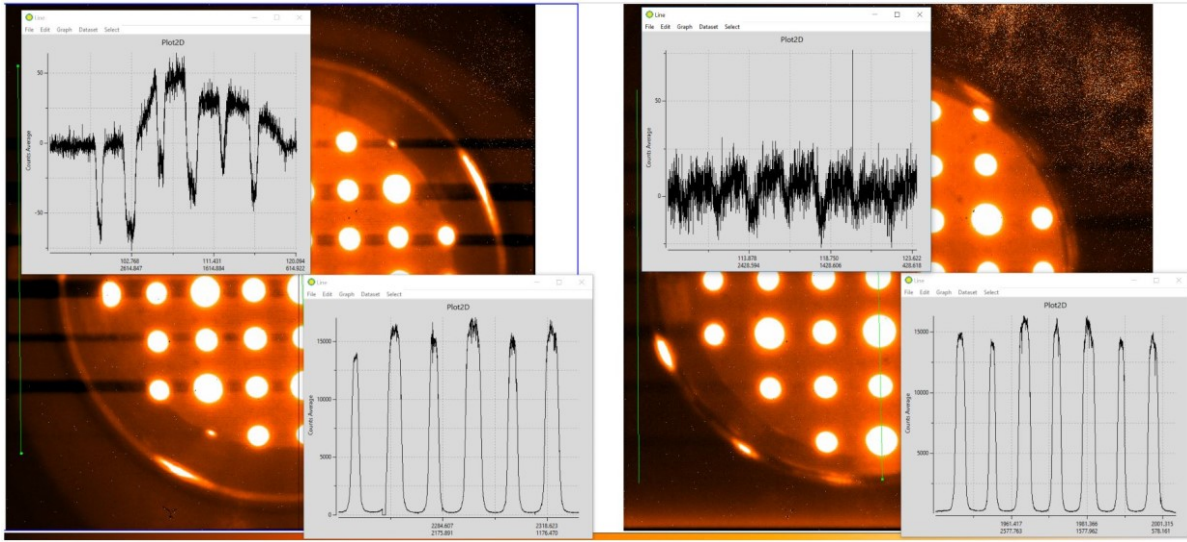

Figure 7, DC crosstalk measurements by projecting a series of spots onto the detector and measuring the droop in the corresponding rows, left for the unbuffered mode and right for the buffered mode.

To test the DC crosstalk or coupling between channels or along a row we projected a series of bright spots onto the detector in both modes. The image in Figure 7 shows the results for the unbuffered on the left and buffered on the right. For each row of spots there is an associated drop in the dc level along the row. The plots associated with each image indicate the signal level of the spots, typically three quarters full well and the droop in the row level in the dark part away from the spots. Typically, the buffered mode reduces this voltage droop by at least a factor of 3 or more along the row compared to the unbuffered mode.

## 5. DARK CURRENT

The dark current was measured from 80K down to 40K. The MOONS instrument will operate at 40K but the MICADO instrument for the ELT will operate at 80K, so it was important to understand operation at both temperatures. At each temperature the dark current was measured from a 10 hour long dark with 120 up the ramp samples. A line fit was made on a per pixel basis using all the samples per pixel, these maps are shown in Figure 8. The dark current was determined for a good pixel region, a hot pixel region and for the full array. The results for each temperature as well as a slightly warmer temperatures are given in Table 1 and the dark bitmaps at 40K and 80K are shown in Figure 8. A large area of hot pixel region can be seen from the dark bitmap at 80K. A few hundred rows at the detector bottom (bond pad side) have higher dark current due to glow at the edge, even at 40K. It is established that this glow is proportional to number of reads (probably fast shift register activity) and is also greater at higher operating temperature. This is reported in more detail later in this paper. In general, the cosmetics improve greatly at 40K, especially the operability of hot pixels (see top right corner of the dark current bitmaps).

Table 1, detector dark current for good pixel region, hot pixel regions and for entire array at different operating temperatures.

| Dark Current | Mean dark generation (e/s/pixel) and (Sigma) | | | |
|---|---|---|---|---|
| | 40K | 80K | 85K | 90K |
| **Good Pixel Region** | 0.004 (0.009) | 0.009 (0.012) | 0.010 (0.015) | 0.014 (0.025) |
| **Hot Pixel Region** | 0.008 (0.032) | 0.159 (0.134) | 0.213 (0.138) | 0.253 (0.137) |
| **Full Array** | 0.010 (0.021) | 0.033 (0.066) | 0.039 (0.077) | 0.046 (0.087) |

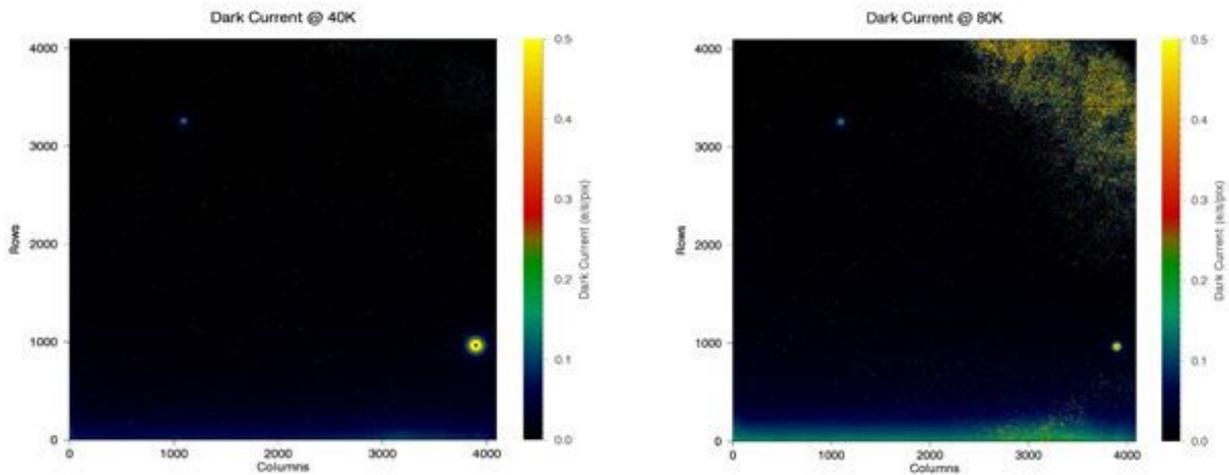

Figure 8, dark current bitmaps of an engineering grade H4RG detector at 40K and 80K.

### 5.1 Noise and noise operability

The detector bias/dark stability and noise operability were determined from two identical bias/dark frame sets acquired up to 4 days apart over a range of detector integration times (6s to 300s) and at different operating temperatures (40K, 80K, 85K and 90K). Two identical frames at a given operating temperature were subtracted from each other and the resultant frame divided by √2 and reference pixel corrected to account for any thermal drifts. The noise and the noise operability statistics for the good region, hot pixel region and for the full array have been computed with a spatial noise limit of ±35e rms. To summarize this, the noise of the good pixel region is slightly improved at 40K compared to 80K, but the noise of the hot pixels is considerably improved at 40K. The noise operability of good pixels remains almost the same at approximately 99.5% from 40K to 90K. However, the noise operability of the hot pixels is reduced by 7.2% going from 40K to 80K and it further reduces by 4.6% from 80K to 85K and another reduction of operability of 5.9% from 85K to 90K. Likewise, the UTR noise statistics of the good pixels remained the same over the 4 day period, whereas for the hot pixels the noise statistics became poorer as the time gap between the data sets increases. That is the pixel dark current levels are less correlated between the different data sets at 80K over 4 days, compared to 40K over the same period.

## 6. PERSISTENCE

We evaluated the persistence seen in the detector as a function of temperature by performing the measurement technique developed at ESO to measure the density of traps with different time constants detailed in Tulloch[5]. At each temperature, the measurement consisted of an LED flash followed by a soak time of 10,000 seconds to allow all traps to come into equilibrium. The detector was reset and then the full array was read out repeatedly at approximately logarithmically spaced intervals for 14 hours to sample the persistence de-trapping curves at time constants between 1 and 100,000 seconds. This measurement was repeated several times at different exposures levels up to and beyond full well to evaluate persistence as a function of exposure time as well as at different diode bias levels.

As was seen in previous persistence measurements at ESO, persistence is a strong function of temperature, and has a complex spatial distribution. Figure 9 shows an example of the spatial distribution of persistence at 40K, 65K, and 80K after approximately 15 minutes of de-trapping as well as the 3 regions of the detector analyzed for the time constant analysis. These regions were selected for their differing behaviour. One appears to be a glue void (bottom left), one is in a good performance region at the center of the array, and one is in a bad region (top right) displaying many hot pixels at higher operation temperatures and overall higher persistence. The images clearly show the evolution of the different behavior at every pixel for the three different temperatures and indicates that the persistence is highest for a temperature of 65K.

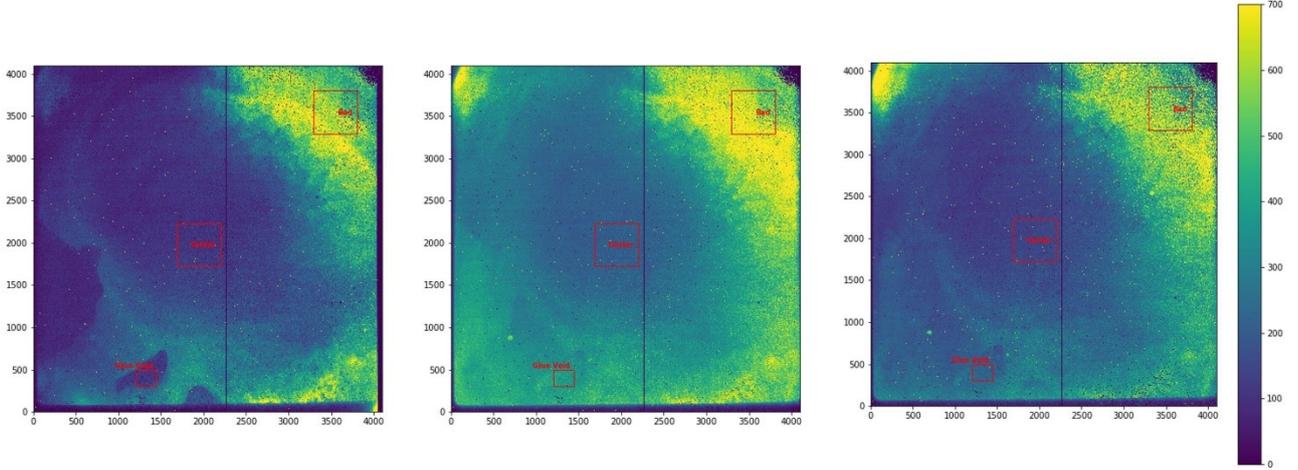

Figure 9, persistence maps (ADU) for the same detector after 830s of de-trapping for 40K (left), 65K (middle) and 80K (right).

Figure 10, which we call the "persistence water-fall diagram" shows the trap density in the 6 time constant bins in these 3 regions at a temperature range from 40K-90K. A feature of all three regions is that there is a peak in trap density that moves to shorter time constants at higher temperatures, and we interpret this as a single trap with a temperature dependent de-trapping time. At temperatures below 50K, this trap is effectively frozen out and remains always full, on timescales of normal astronomical exposures. At temperatures above 75K, the thermal energy is sufficient for this trap to de-trap after detector reset on timescales so fast that most traps are empty by the time we read the first frame after detector reset.

If we take the peak in the time constant density at each temperature as the time constant of the trap, and assume that the trap is de-trapping according to SRH (Shockley Read Hall), the following equations 1 and 2 allow us to fit the data with two free parameters, $E_{trap}$, the activation energy and $\sigma_{trap}$ the capture cross section:

$$\tau_{detrap} = \frac{1}{AT^2} e^{-\frac{E_{trap}}{kT}} \qquad (1)$$

$$A = 4\sqrt{6}\ \pi^{3/2} \sigma_{trap} \frac{k^2 m_{eff}}{h^3} \qquad (2)$$

where $m_{eff}$ is the effective mass of the charge carrier. Figure 11 shows the results of the fit. It should be noted that our measurements are only sensitive to time constants between 1 and 10^6 seconds, so at times shorter or longer than this range we only have upper and lower limits on the $\tau_{detrap}$ respectively. Additionally, the emission time constant can be affected by diode reverse bias, and our data is insufficient to pinpoint the depth in the depletion region of these traps. We can therefore interpret the fit value of $E_{trap}$ as an upper limit on the trap activation energy.

The best fit results are $E_{trap}$ = |0.13 +/-| 0.01 eV, with A and therefore $\sigma_{trap}$ essentially unconstrained, however the fit is not very sensitive to A. For this detector $E_{gap}$ ~0.5 eV, putting this trap at $E_{gap}/4$. Trap energies measured in passivated HgCdTe using Deep Level Transient Spectroscopy (DLTS) in the literature cover a broad range of energies.[6] This method of

measuring persistence by resetting the detector after saturation and measuring the de-trapped charges is similar in concept to DLTS, but is sensitive to a very different range of time constants.

The presence of this 0.13 eV trap in the detector indicates that to minimize detector persistence for astronomical exposures between a few seconds and up to an hour in length, the detector should be operated at either less than 50 K or greater than 75 K. We also note that at higher temperatures (> 80 K), especially in the bad region of the detector, there are hints of another higher energy trap coming in at long time constants that causes increasing persistence again at temperatures >80K. However, we do not have enough data at higher temperatures to fit for this trap energy. For minimum persistence we therefore recommend operating either at ≤50 K or at ≥80 K. More details of this measurement and modeling of the trap behavior using PyXel, our detector modelling collaboration with ESA, will be provided in a forthcoming paper.

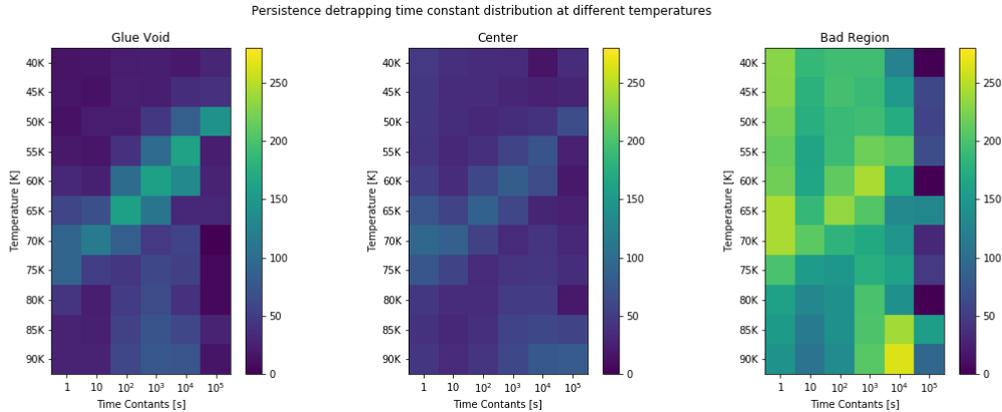

Figure 10, waterfall diagram of trap density in the different time constant bins in three sub-regions of the detector.

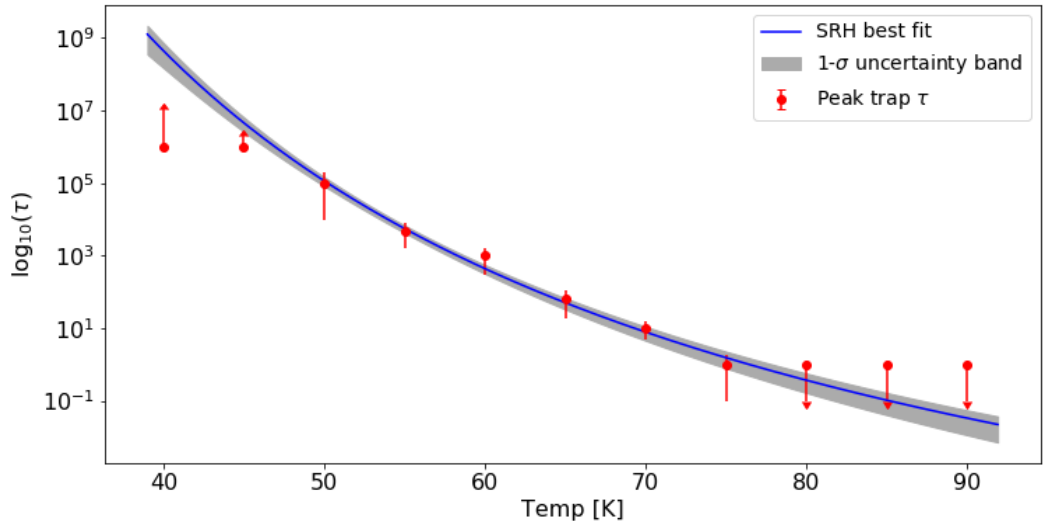

Figure 11, fit of the SRH de-trapping model to the peak time constants observed at each temperature.

## 7. INTERPIXEL CAPACITANCE

A robust method of multiple column / row resets is used to measure the inter-pixel capacitance (IPC) across the H4RG array. The method[7] involves taking two identical CDS flats, one with multiple single column / row resets and another flat without resets (normal flat). The CDS flat with column/row resets is subtracted from the normal flat. In the resultant frame, the counts in the neighbouring columns/rows are normalised with the counts in the reset columns/rows. This results directly in the amount of charge (in %) coupled into the columns/rows which are at reset level from the immediate neighbouring columns/rows. An average of multiple CDS flats is used to improve the measurement accuracy. This is similar to the single pixel reset method but extended to multiple columns/rows to measure IPC across the array quickly. Due to memory mapping limitations of the clock patterns in our controller, only 6 static columns/rows could be defined at once and hence the IPC in the column and row directions separately. The flats are generated using a LED and are not uniform across the array.

## 7.1 IPC in row direction

In the unbuffered output mode, the pixel frequency response is highly dependent on the current through the unit cell. Insufficient current through the unit cell results in pixel lag in the direction of the readout as the output takes relatively longer to settle to its final signal level. This results in non-symmetric coupling measurements along the row direction. For a 100kHz pixel rate, it was found that a VBISGATE of 1.7V results in no pixel lag and hence a symmetric coupling to the neighboring pixels in the row direction. Figure 12 shows a slight gradient from bottom of the array (close to bond pads) to the top (away from the bond pads) in the IPC. The larger spread in the data towards the top is due to the less signal as the LED flats are not uniform.

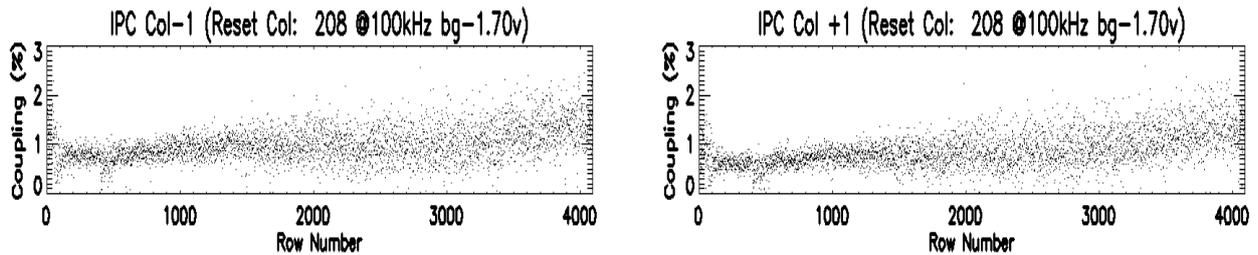

Figure 12, a symmetric coupling to the neighboring columns with VBIASGATE at 1.7V shows the pixel frequency response is good at 100kHz in unbuffered output mode.

In order to confirm the variation in IPC is not due to the signal gradient, the coupling is measured from 4 different regions of the array separately. The average signal level in these regions is kept about the same level (about 50ke) by adjusting the LED exposures for each region accordingly. The coupling in the row direction shows a gradual increase from the bottom to the top regions, Figure 13, confirming that the IPC gradient seen previously. The gradient (from 0.8% to 1.4%) remains the same irrespective of the vertical readout direction.

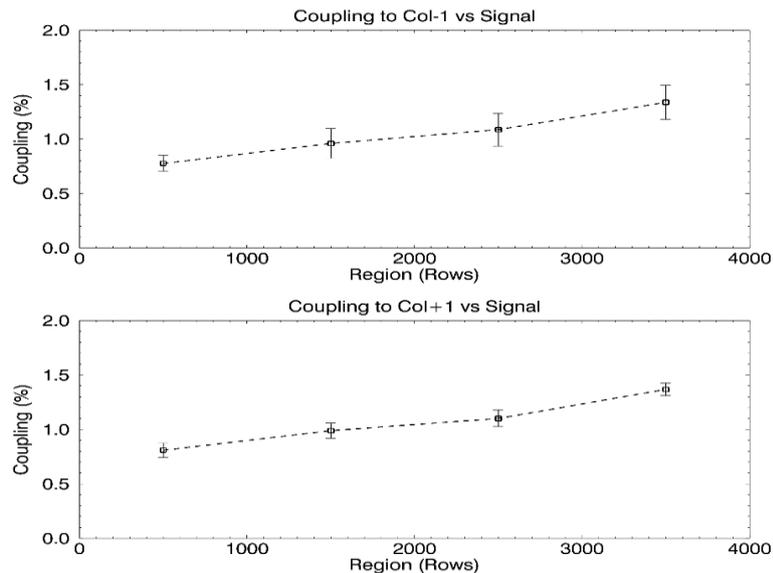

Figure 13, IPC variation from bottom, close to bond pads to top of the array.

## 7.2 IPC in column direction

Similarly, a set of row resets across the array are used in order to investigate the coupling in the column direction. As expected, the pixel frequency has no effect in this case on the coupling to the pixels in the column direction and shows an IPC of 0.5% to the neighbouring rows. However, a strange ripple (triangular shape) is seen in the coupling to the row above (row+1) the reset row as shown in Figure 14. The average coupling seems to be about 0.4% to 0.6% in the column direction.

The ripple seen in the row above the reset row is horizontal readout direction dependent. That is, when the direction is changed from the default (→ ←) to the (→ →) readout direction, the shape of the ripple changes from triangular to sawtooth shape. That is the first pixel read out in a channel has less coupling compared to the last pixel read out in that channel. Strangely, this is seen only in the row above the reset row (row+1). The ripple always appears in a row read out after the reset row in either vertical readout direction, indicating the coupling is modulated by the readout circuits.

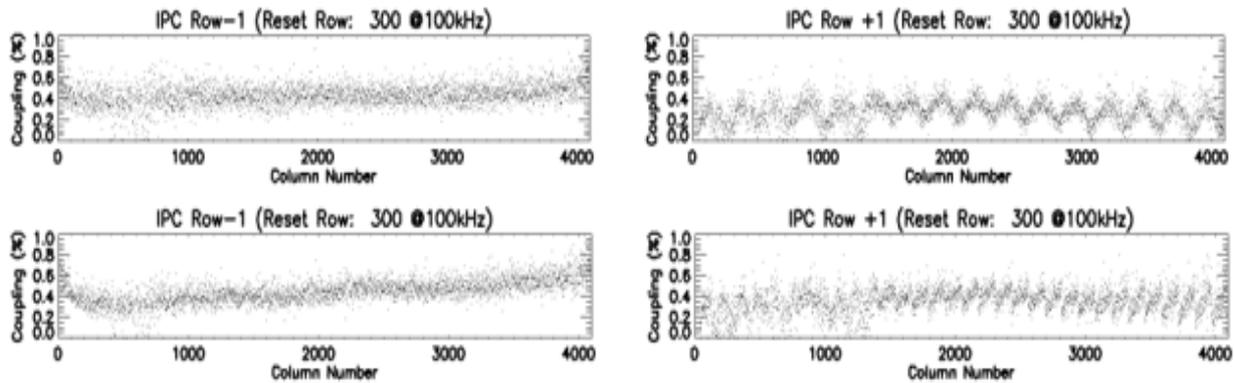

Figure 14, IPC along the column direction. Top half – A triangular shape ripple in the coupling to the row above the reset row in the default horizontal readout direction (→ ←). Bottom half – The ripple changes to sawtooth shape when the readout direction is changed to (→ →). Note the array is operated in 32-output channel mode.

## 7.3 Signal dependency of coupling

The measurements are repeated at 6 different illumination levels in a region at the top the array to see any dependency of IPC on the signal level. This testing indicates that there is almost no dependency of the coupling along the row direction or in the column direction on the signal level.

# 8. UP-THE-RAMP NOISE PERFORMANCE

Data has already been presented previously indicating that the buffered mode gives the same noise performance as the unbuffered mode but wins in terms of signal-to-noise performance per unit time for exposures less than 100 seconds because of the faster frame rate. However, the data also hinted that for long exposures there was a slight trend for increased noise compared to the unbuffered mode. It was thought that this was due to a charge injection per read issue rather than inherently higher noise, the unbuffered mode using less reads up-the-ramp because the frame time was three times slower. To test this hypothesis the detector was operated in an UTR mode using exposures times of 900 seconds where in one case the maximum number of reads, 525, were made in that total time and then this was compared to a similar exposure time but limited to 128 reads in the same time. The results of this test are clearly shown in Figure 15.

The data was produced by increasing the exposure time by 10 seconds up to approximately 900 seconds and then the maximum number of reads was inserted into that time or else only a maximum of 128 reads in total were used, in this case the time between reads was incremented accordingly. Two similar images were taken at the same exposure time and these were subtracted from each other, divided by √2 and then gaussian fitted to determine the noise over a large region of the detector. For both plots the red line indicates the data for the fixed 128 reads and the blue plot is for the 525 reads. The noise is the same in both cases up to 128 reads but then shows that for more reads the noise is increasing. The middle graph replots the same data but now as a function of exposure time, in the case of the 128 reads, a longer time between each read is implemented. This plot clearly indicates that there is some glow associated with each read which adds to the overall noise as seen in the blue plot with the much higher number of reads. The bottom plot shows the signal generated for each read, the blue plot is again for the case of 525 reads and the red plot is for 128 reads. This plot clearly indicates that there is definitely charge injection per read and of the order of approximately 0.03 e-/read. This level of charge injection is enough to reduce the UTR noise performance for long integrations if the maximum number of reads are crammed into the integration. Therefore, for the MOONS instrument we will limit the number of reads to 128 or less.

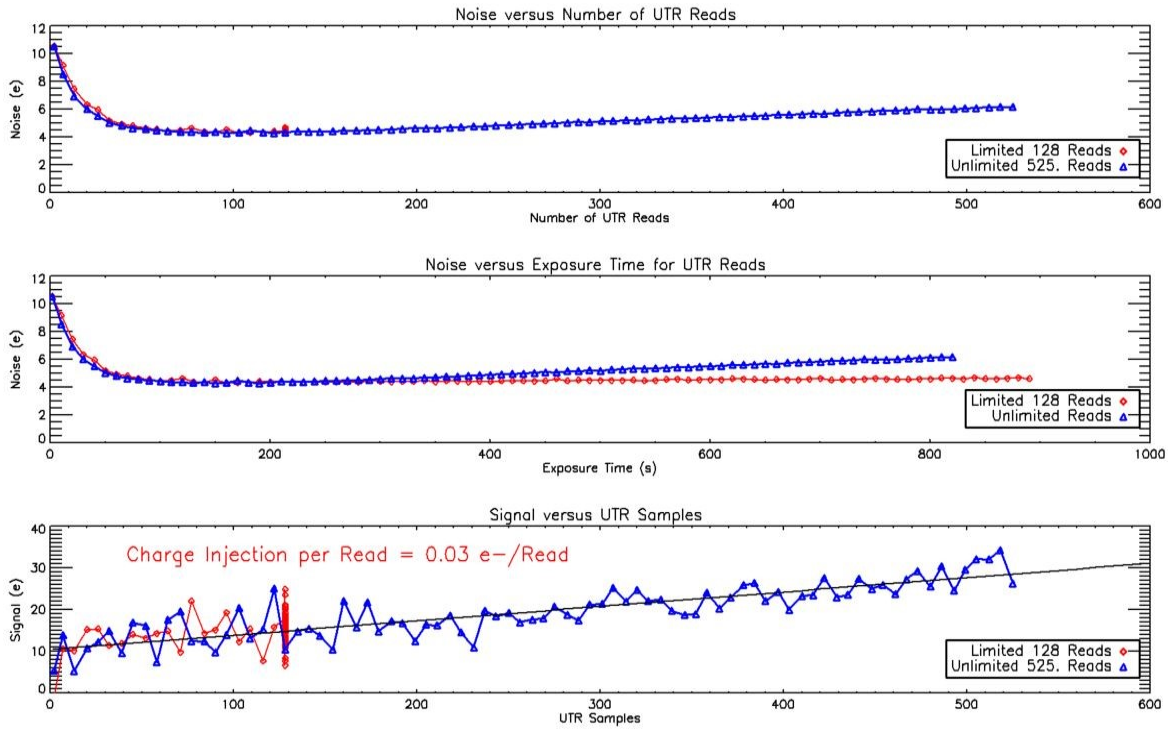

Figure 15, noise versus number of Up-The-Ramp non-destructive reads for 525 reads versus 128 reads in the same exposure time (top), versus actual exposure time, same data (middle) and the signal per read (bottom).

## 8.1 Detector read glow/charge injection

As previously discussed in the UTR performance section, glow per read is evident in the detector. The number quoted was for a cosmetically clean area in the centre of the detector. However, it is obvious from Figure 16 that the there is even higher glow at the bottom of the detector near the readout circuitry. A line plot is shown from the bottom to the middle of the detector with clearly more signal at the bottom than the middle. At first it was thought that this excess signal at the bottom was due to a light leak in our test facility or from glow in a preamplifier which sites just below the detector. However, we could not trace any source of light leakage in the cryostat and we also reduced the power to the preamp without changing the glow pattern. We revisited the datapacks supplied from TIS and saw that all the delivered devices had similar glow at the bottom which they also attributed to a light leak. However further testing using the method already described for UTR performance testing leads us to the conclusion that there is more charge injection or glow produced at the bottom of the detector than in the rest of the detector. This is clearly seen in Figure 17 which is the same plot as previous but now for the bottom region of the detector. It is clear that the glow level is approximately 0.65 e-/pixel/read, more than an order of magnitude higher than for the centre of the detector. The middle plot is the most interesting one which clearly indicates that the glow is associated with the reading of the detector. Here the blue plot is for reading UTR with the maximum number of reads of 525, the signal clearly increases in linear fashion per read. However, the red plot is where the number of reads has been limited to 128, for increasing exposure time the time between each read is increased accordingly. So, for less than 128 reads the signal increases linearly then at 128 it abruptly stops because there is always the same number of reads but only increasing exposure time. If the glow were from an external source in the cryostat then this plot would continue to increase. The plot also indicates that 1/f noise is minimal in our system such that it has no effect on the noise at least for 900 second exposures or less.

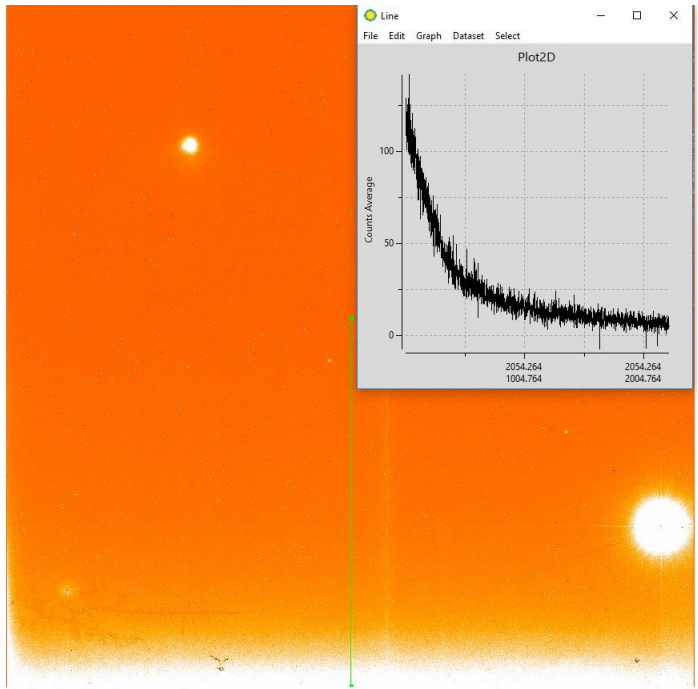

Figure 16, UTR ramp image showing excess glow at bottom of detector near the bond wires.

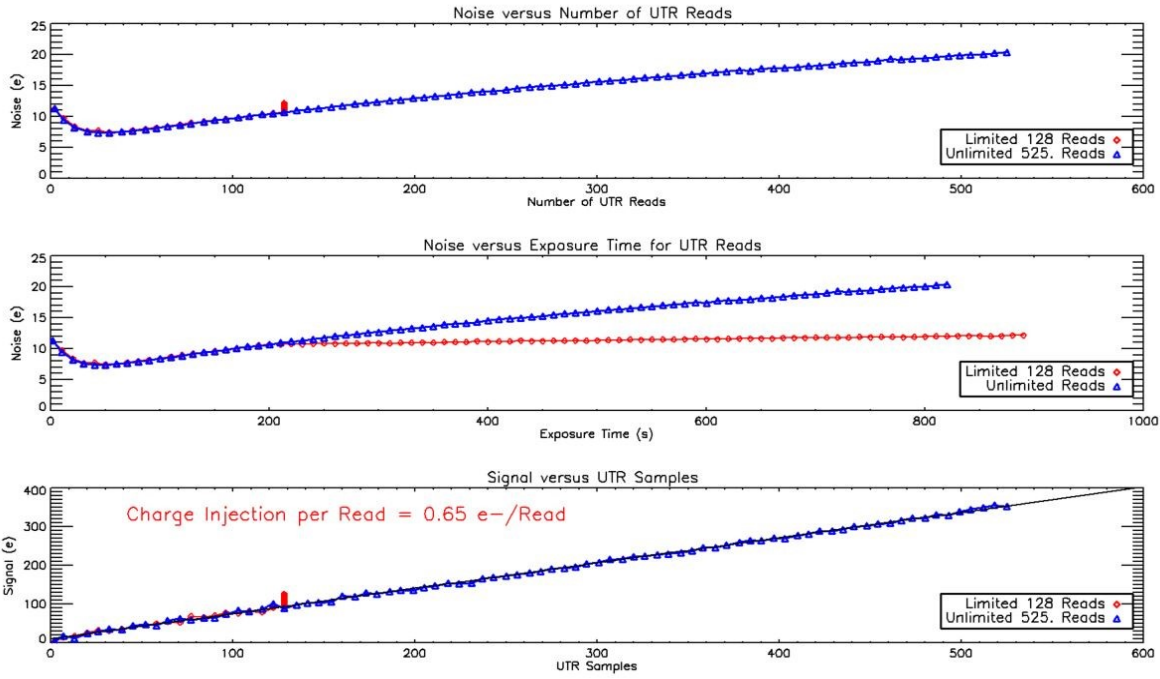

Figure 17, noise versus number of Up-The-Ramp non-destructive reads for 525 reads versus 128 reads in the same exposure time (top), versus actual exposure time, same data (middle) and the signal per read (bottom), all data for the bottom region of the detector.

# 9. READ NOISE, QUANTUM EFFICIENCY AND OTHER MEASUREMENTS

The read noise has already been measured with the Up-The-Ramp mode as indicated earlier. For a CDS it is typically 13e- rms and this reduces to ~4.5e- rms for 128 non-destructive reads. For Fowler sampling we achieve less than 3e- rms. As a follow up to this analysis we also did a temporal power spectral analysis for a CDS readout. This is determined by taking many thousands (2048) of CDS frames and then doing a power spectral density analysis for a single pixel. This calculation is then repeated for many other pixels (1000 x 600) and the resulting data is averaged. This average power spectral density is given in Figure 18. The detector read noise can be determined from the area below the plot, indicating it is approximately 13e- rms which is in very good agreed with our spatial measurements of noise. The plot also indicates that our detector, preamplifier, cable and control electronics setup does not suffer from 1/f noise, at least not for typical exposure times used for the MOONS instrument which is consistent with our previous measurements.

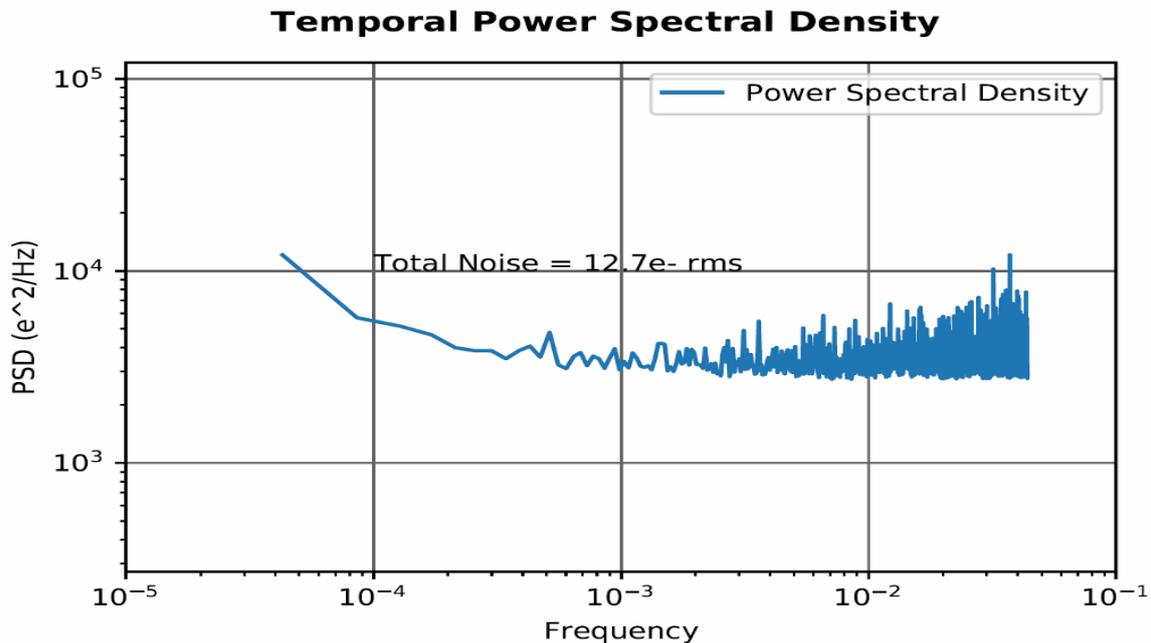

Figure 18, temporal power spectral density plot for CDS readout, calculated from the average of many pixels.

## 9.1 Linearity

The detector linearity was determined by flashing an LED inside our cryostat and then plotting the LED flash time (frame number here) versus signal. A second order polynomial was then fitted to the data and the difference between the polynomial and the real data was determined. This is shown in Figure 19 which indicates that the linearity is good to +/- 1% over at least two thirds of the dynamic range of the detector. At 0.25V bias we reach a full well of greater than 70ke-, assuming that full well is where the signal is no longer governed by Poisson statistics.

## 9.2 Quantum Efficiency measurements

Our test facility was developed to allow for rapid operation and testing of the detectors. It is a simple commercial cryostat with a cold pinhole, filter and window arrangement which is not ideal for doing accurate QE measurements. However, we have tested for QE using an external blackbody and have results for both the astronomical J and H bands. We typically measure detector QEs of approximately 90%, with large error bars dues to some internal reflections inside our cryostat. These measurement values are expected values for the detector, so we have confidence that the detector QE is as expected. At a later stage our new test facility will come online and this will allow for more accurate QE measurements as well as scanning with a monochromator to measure the detector cut-off wavelength.

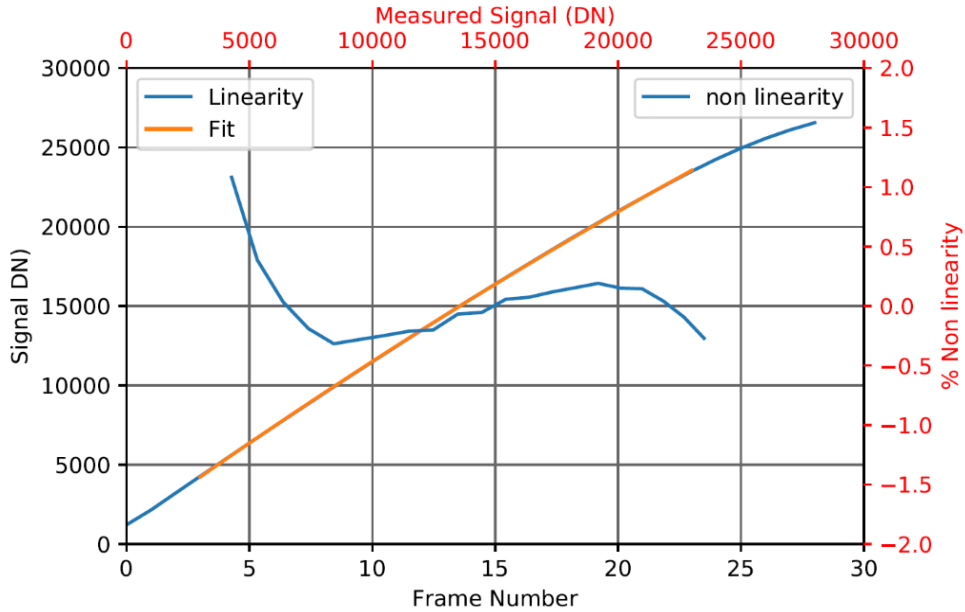

Figure 19, linearity of detector, the curved plot is the non-linearity which is the second order difference between the real data and a 2nd order polynomial.

## 10. DETECTOR OPERATION TIPS AND TRICKS

The Hawaii-RG detectors can be operated in number of ways by programming the various on-chip registers. The operations are well described in the operating manual. However, there are some register bits that are not well explained for their intended use. One of them is the **current boost enable bit** intended to be used with the slow un-buffered operation.

### 10.1 Current boost operation

In the slow unbuffered mode, the engineering detector is quite unstable after a power on and takes up to half an hour to be able to get consistent frames. The current boost enable option in the output mode register eliminates this problem and results in stable frames almost immediately. Figure 20 show the difference with and without the current boost enable operation soon after power-on of the detector. Further to this even though a typical CDS images indicates that the detector

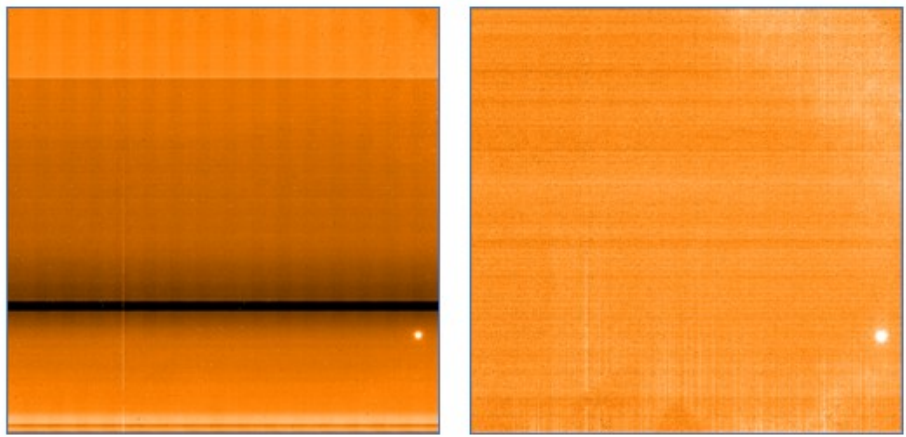

Figure 20, Detector stability improves quickly after power on with current boost enable. Left – Normal operation without current boost enabled, Right – with current boost enabled.

settles quickly, recent analysis has shown that even with the current boost enabled, after power up it is better to wait for at least 1 hour of settling time if the user wants to achieve the lowest noise performance using multiple non-destructive reads

**10.2 Photoemission defect suppression**

This engineering grade detector has a bright photo-emission defect (PED) at the bottom right corner. When the column that the PED is located in is disabled then the PED glows even stronger. However, when the row in which the PED is located is skipped, the glow can be suppressed. In order to preserve the correct array size and the frame time, the skipped row data is replaced by repeated sampling of the reference pixels at the beginning of that row. Figure 21 shows the suppression of the PED when the corresponding row in which it is located is skipped, as described.

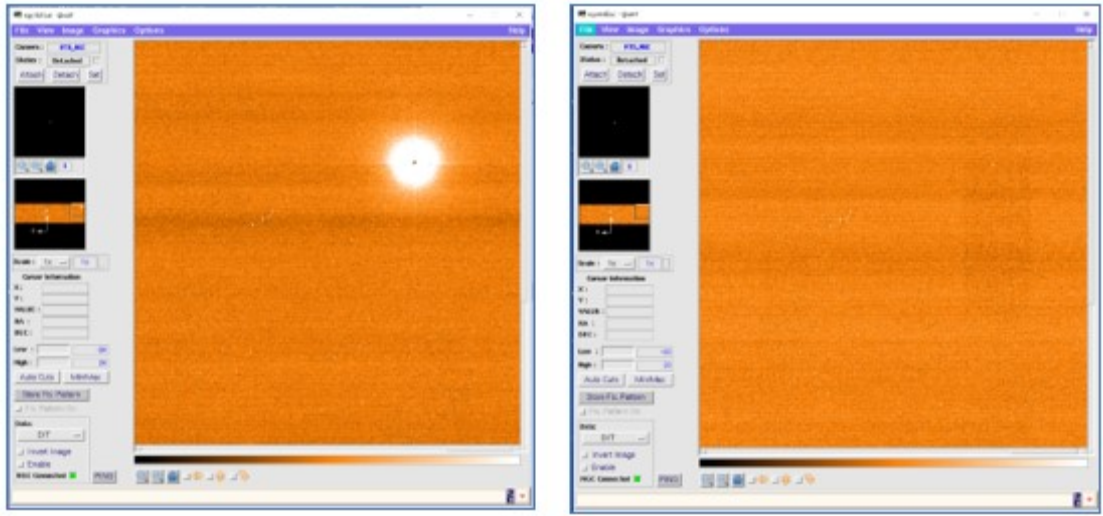

Figure 21, Photoemission defect is suppressed by skipping the row in which it is located. Left – Photoemission defect in normal operation. Right – PED suppressed when the row is skipped.

**10.3 Register 1 programming effect**

It has clearly been observed that not writing to register 1 of the detector register set diminishes the performance of the detector. Even writing only default values to this register gives better performance than not writing to the register. Register 1 with binary address 0001 is the Reference Pixel Row Option register. It allows for programming the functionality of the reference rows and columns. Typically, if the detector is used in its default mode then the capacitor reference pixel parallel is selected. Being lazy programmers then we typically only write to the registers if and only when we need to change their status. We do not typically write default values to the registers. However, as will be explained it is actually necessary to at least to write the default values to the Register 1. To determine the effect of programming the register or leaving it in its default state with no programming, a Photon Transfer Curve was calcuated for each mode of operation. The test was repeated for CDS with true global reset, CDS with Pseudo Global Reset, where the global reset is executed by using the row reset for all rows before readout and finally for Row Reset Read where a row is reset then read.

For each PTC measurement, a plot of Variance versus Signal was made, together with a linearity plot, a plot of calculated gain versus signal level and finally a histogram of calculated gain for low signal level. It should be noted that the problem is only with Register 1. Only the Row Reset Read data is shown here, however the same issue is seen with all reset modes. Figure 22 shows the plots for the case where the Register 1 is not written to. Figure 23 is the same data again but after writing the default values to the same register. The configuration and test setup was the same in both instances. It is obvious from all the plots that writing to Register 1 increases the transimpedance conversion gain in uV/e by approximately 20%, with a slight reduction in read noise and an increase in the detector full well from less than 40ke- to more than 60ke-. It is also true that writing default values to the register gives higher full well than not writing to the register. This is a strange effect and as yet not understood why not writing to the register can affect the full well and gain in such a way.

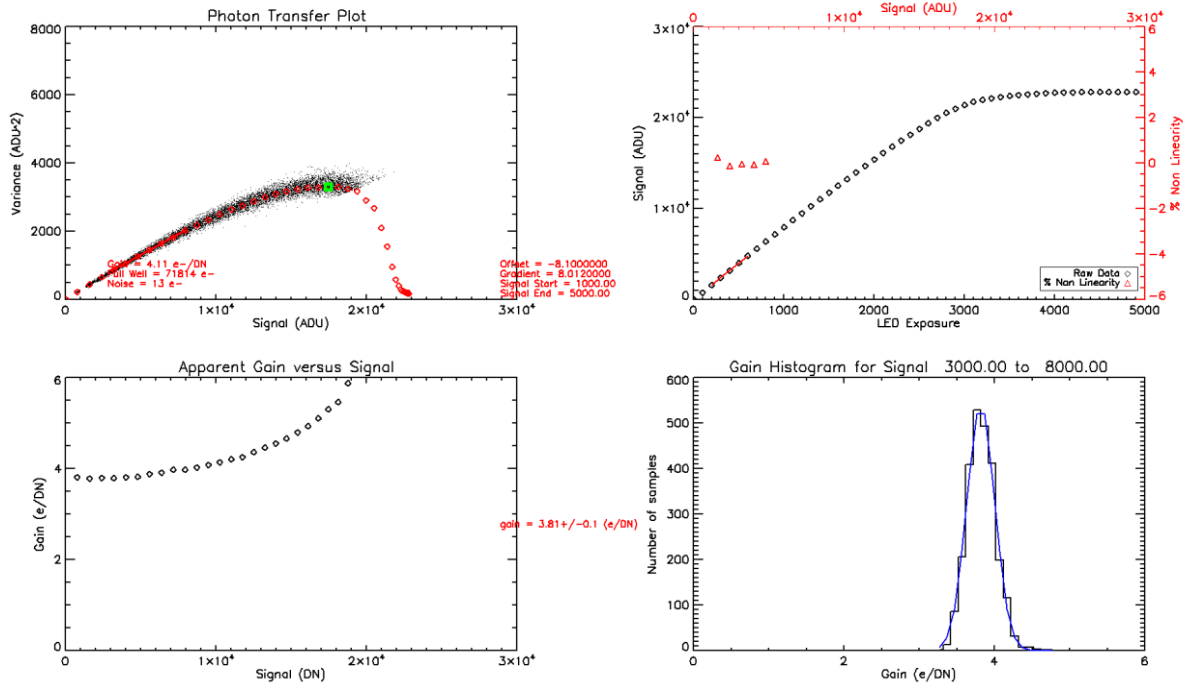

Figure 22, Photon Transfer, linearity, apparent gain versus signal level and average gain histogram when Register 1 is not written to.

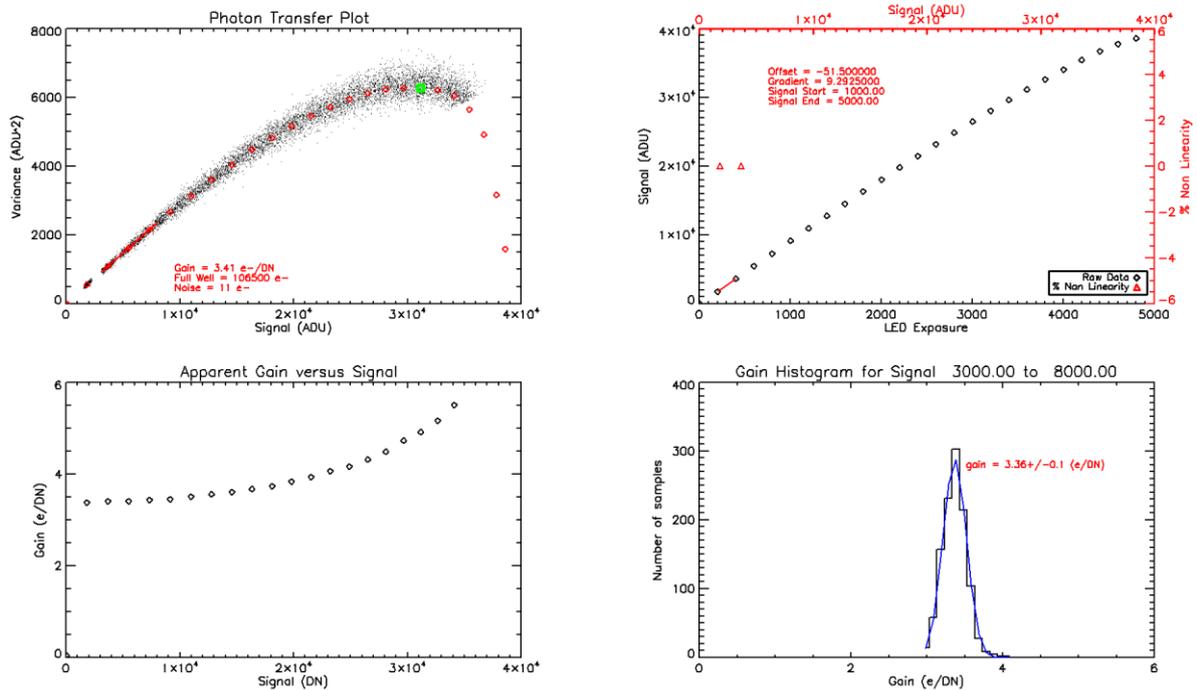

Figure 23, Photon Transfer, linearity, apparent gain versus signal level and average gain histogram when Register 1 has the default values written to it.

## 10.4 Apparent effect of global reset versus row reset on Inter-pixel capacitance

Depending on how the detector is reset can have an effect on the apparent inter-pixel capacitance shape. Typically for a global reset the inter-pixel capacitance is seen as coupling to any one pixel's nearest neighbour, in a reasonably symmetrical fashion and for H4RG detector at a level of typically less than 1% of the hot pixel signal level, to the 4 nearest cardinal neighbouring pixels. Thi is clearly seen for hot spots on the detector, as indicated in Figure 24 where the image to the right is the average of many hundreds of hot pixels and the plot on the left is the average measured coupling as a percentage of these hot spots to their nearest neighbours. We measure less than 1% coupling to nearest neighbours.

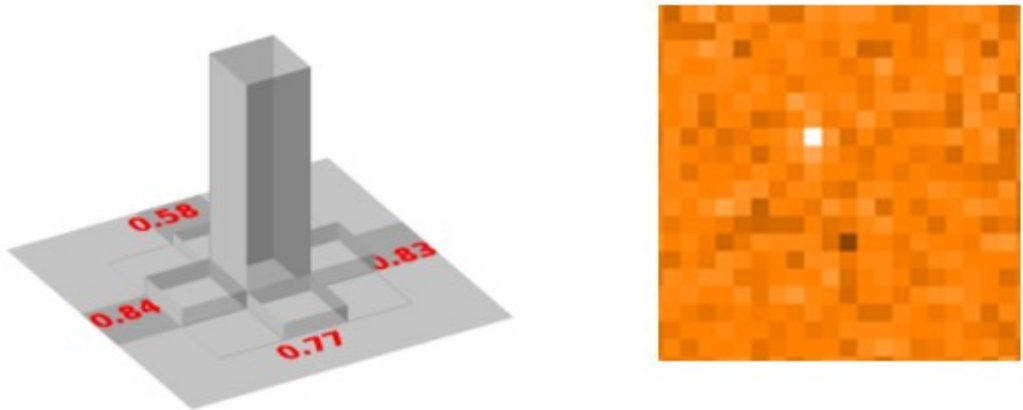

Figure 24, averaged hot spots on the detector (right) and a profile of the averaged hot spot as a function of percentage signal coupled from hot spot to its nearest neighbouring pixels (left) for a global reset mode.

In the Row-Reset-Read per row mode then this standard "+" shaped IPC pixel becomes an inverted "T"shape pixel instead. This is clearly seen in Figure 25, where again the hot spots are averaged across the detector. In this case a clear upside down "T" shaped hot spot is seen where the apparent coupling to the pixel in the row which has been reset seems to have no coupling. The coupling to the other three pixels is as expected.

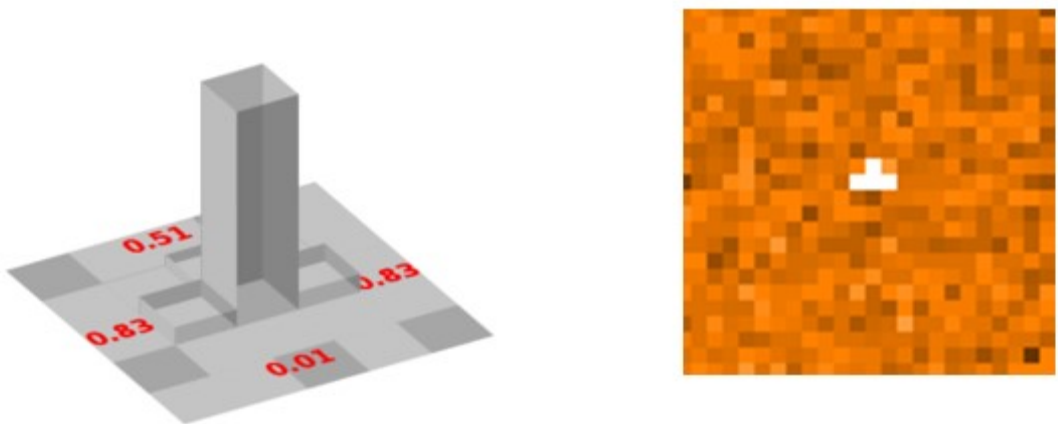

Figure 25, averaged hot spots on the detector (right) and a profile of the averaged hot spot as a function of percentage signal coupled from hot spot to its nearest neighbouring pixels (left) for row reset read mode.

The setup shown was for a Row-Reset-Read for the first frame followed by a simple Row-Read for the second frame of a CDS sequence. However we have also tested a read scheme with Row-Reset-Read for the first frame and Row-Read-Reset for the second frame and in this case the coupling looks like a "-"rather than an inverted "T" or "+"which is what is typically expected. Our theory is that the coupling is of course instantaneous between hot pixels and their nearest neighbours and before these rows are read out. Then the row before the hot pixel is reset and read out, resetting the coupling which had occurred in that row from the hot pixel in the next row. The same thing occurs in the reverse sequence of a Row-Read-Reset.

**10.5 Enhanced Clocking read mode**

Enhanced clocking is another possible clocking mode which can be used with the H*RG family of detectors but has never been used before at ESO. This mode allows for reset read on a pixel by pixel basis rather than using the more typical global reset which resets all pixels at the same time or row reset which resets a complete row at the same time. TIS had stated that there might be some cosmetic advantages to using this enhanced clocking mode. The mode is enabled by setting the appropriate bit in the "NORMAL MODE" register. The clock waveforms then also need to be changed as per the detector user manual.

The following image in Figure 26 is a typical uncorrelated image showing one of the main problems with this read mode where the first column for every output is approximately 20k counts above all the other columns for the particular outputs. For a CDS image this would not be seen because the columns subtract out precisely, however it does mean that the dynamic range of the first column in every output is severely limited.

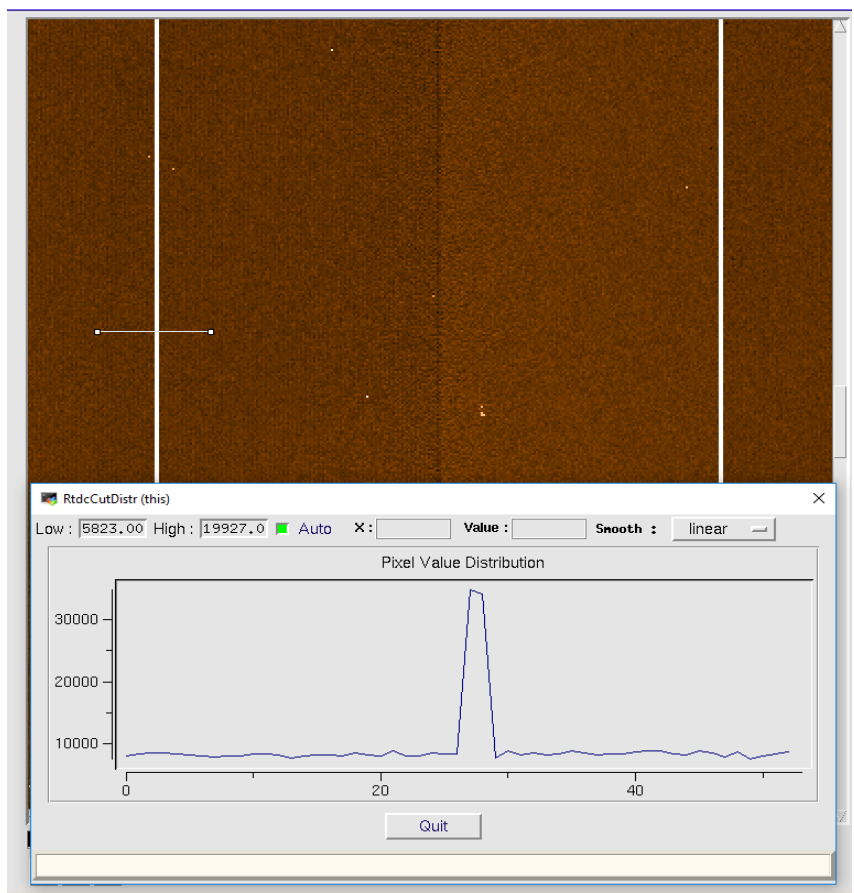

Figure 26, uncorrelated read image using enhanced pixel clocking indicating problem seen with first column.

## 11. CONCLUSIONS

The first H4RG detector with 15 µm pixels for the MOONS project has been extensively tested. This is an Engineering Grade device with mostly cosmetic issues in one corner. However, in most other aspects of astronomical performance it is a good device. We find that operation at 40K improves the cosmetics of the device compared to 80K, however there is little or no difference in other aspects of operating the device at the two different temperatures in terms of dark current, persistence, and stability etc. We do measure some charge injection per read which seems to be higher than for H2RG detectors and which means that we have to limit the number of non-destructive reads in an Up-The-Ramp exposure to give the lowest noise performance. We still achieve ~4.5e- rms in such a mode which is an excellent result, this equates to less than 3e- rms for the Fowler read mode. We have measured the persistence in this detector from 40K to 80K and show that persistence is indeed best at the lowest temperatures but is almost as good at the higher temperature of 80K. However, at temperatures in between from 55K to 70K then the measured persistence is highest, due to the time constant of the main trap being of the order of a typical astronomical exposure. The detector has been operated in unbuffered with 32 channel readout mode where it was difficult to achieve 100 kpixel/s read rates without increasing markedly the unit cell current. The detector has also been operated in 64 channel buffered readout mode which will now be our default setup for all future instruments using the H*RG family of detector. The buffered mode allows us to read the detector at higher pixel rates with almost a flat noise response for exposures of 1000s or less whilst also almost eliminating the electrical crosstalk between the readout channels. A new, low power, low temperature drift op-amp have been successfully tested at cryogenic temperatures and which showed a superior performance compared to the currently used op-amps for slow speeds. This has been implemented in a very complex flexi-rigid PCB for the MOONS project to ensure that the cold electronics do not vignette the optical beam in this instrument.